\documentclass[12pt]{article}

\usepackage{amssymb}
\usepackage{amsmath}
\usepackage {pstricks}
\usepackage{graphicx,color}
\usepackage{mathrsfs}
\newcommand{\al}{\alpha}
\newcommand{\be}{\beta}
\newcommand{\mcN}{\mathcal{N}}

\numberwithin{equation}{section}
\usepackage{caption} 
\newcommand{\bi}{\bfseries\itshape}

\begin{document}

%
\begin{titlepage}
\begin{flushright}
{RIKEN-MP-70}\\
{March, 2013 }
\end{flushright}
\vspace{1.0cm}
\begin{center}
{\Large \bf 
\textsl{}
\textsc{Holomorphic Blocks for\\
\vspace{0.3cm}3d Non-abelian Partition Functions
}}
\vskip1.0cm
{\large Masato Taki
}
\vskip 1.0em
{\it
Mathematical Physics Lab., RIKEN Nishina Center,\\ 
Saitama 351-0198, Japan
}%
\\
\vspace{0.5cm}
{\tt taki@riken.jp}
\end{center}
\vskip1.3cm

\begin{abstract}
The most recent studies on 
the supersymmetric localization reveal many non-trivial features of
supersymmetric field theories in diverse dimensions,
and 3d gauge theory provides a typical example.
It was conjectured that
the index and the partition function of a 3d $\mcN=2$ theory
are constructed from a single component:
the holomorphic block.
We prove this conjecture for non-abelian gauge theories
by computing exactly the 3d partition functions and holomorphic blocks.
\end{abstract}
\end{titlepage}


\renewcommand{\thefootnote}{\arabic{footnote}} \setcounter{footnote}{0}


\section{Introduction}

The pioneer work by Pestun  \cite{Pestun:2007rz} on the partition function
of four-dimensional (4d) $\mcN=2$ theories 
has served as a trigger to great progress on localization computation of 
supersymmetric gauge theories
in diverse dimensions and on various manifolds \cite{Festuccia:2011ws}.
Localization
of three-dimensional (3d) theories is a focus of recent attention.
Kapustin, Willett, and Yaakov \cite{Kapustin:2010xq,Kapustin:2010mh}
extended Pestun's idea to gauge theories on $S^3$,
and they obtained matrix model representations for the supersymmetric partition functions
of these theories.
We can solve these matrix models in large-$N$ limit,
for instance the ABJM partition function was computed by Drukker, Marino, and Putrov 
\cite{Drukker:2010nc}.
They found that the free energy of the ABJM theory
actually shows the $N^{3/2}$-scaling behavior which 
had been suggested by the AdS/CFT argument.
This result is a typical example of the power of the localization approach.
 
The efficiency of localization reaches beyond large-$N$ approximation.
The matrix models for
partition functions of $\mcN=2$ gauge theories on $S^3$
was derived in \cite{Jafferis:2010un,Hama:2010av}.
The integrant of this matrix model consists of 
a complicated combination of double-sine functions,
and it looks hard on first glance to evaluate it exactly.
In \cite{Benvenuti:2011ga,Pasquetti:2011fj}, however,
the authors succeeded to solve these matrix models exactly.
In particular the partition functions
of 3d $\mcN=2$ $U(1)$ theories computed in \cite{Pasquetti:2011fj}
show the following factorization property:
\begin{align}
\label{Pasquetti}
Z^{U(1)}[S^3]
=\sum_i Z_{\textrm{vort}}^{(i)}\,\widetilde{Z}_{\textrm{anti-vort}}^{(i)}.
\end{align}
Here $Z_{\textrm{vort}}$ and $\widetilde{Z}_{\textrm{anti-vort}}$ are the 
K-theoretic vortex/antivortex partition function 
\cite{Shadchin:2006yz,Dimofte:2010tz} on $S^1\times \mathbb{R}^2$.
The summation is taken over the supersymmetric ground states
which specify the vortex sector.
This factorization into vortexes is 3d analogue of Pestun's expression
\begin{align}
\label{Pestun}
Z^{U(1)}[S^4]
=\int da\, Z_{\textrm{inst}}(a)\,\widetilde{Z}_{\textrm{anti-inst}}(a).
\end{align}
In this 4d case, ground states are labeled by the continuous moduli parameter $a$,
so we take the integral over it after combining the contributions from
instantons and anti-instantons.
3d factorization is therefore
expected to originate from the localization 
after changing the way of it\footnote{
The factorization of 2d theories was shown
along the line \cite{Doroud:2012xw,Benini:2012ui}}.

In this article we prove the factorization of this type
actually occurs in non-abelian gauge theories.
The matrix model for a non-abelian theory
involves a complicated interaction,
and so it is not easy to compute it straightforwardly.
We therefore employ the Cauchy formula\footnote{
This idea was suggested in \cite{Pasquetti:2011fj}.}
and we resolve the problem into that of abelian theory.
We fond that
the factorized partition function is consistent with
the vortex/antivortex partition functions for the corresponding non-abelian theory.
Our result strongly supports the conjecture \cite{Beem:2012mb} 
on the factorization of generic 3d $\mcN=2$ theories.

This article is organized as follows. 
In Section 2, 
we review the factorization of supersymmetric partition functions and superconformal indexes
of 3d $\mcN=2$ gauge theories.
In Section 3, we compute exactly the partition functions 
of non-abelian gauge theories 
based on the matrix model representation coming from localization. 
We then find that the partition functions are actually factorized into
the holomorphic blocks.
The topological string interpretation of these holomorphic blocks
is given in Section 4.
Section 5 is devoted to discussions of our results and future directioins.


\section{3d partition functions and factorization}

In this section we provide a review of
localization and the resulting factorization
of the partition function and 
the superconformal index of a 3d gauge theory with $\mcN=2$ supersymmetry.
The factorization is only conjecture yet for generic $\mcN=2$ theories,
however, there exists a nice geometric interpretation of this phenomenon.

The partition functions of 3d theories
were calculated with the help of supersymmetric localization.
The path integral for a theory on squashed three-sphere is  
\begin{align}
Z
=\int \mathcal{D}\Psi
e^{-S[{S^3_b}]-t\{Q, V\}},
\end{align}
For suitable choice of the scalar supercharge $Q$ 
and the deformation action $V$,
we can calculate it exactly in the limit $t\to\infty$ \cite{Hama:2011ea,Imamura:2011wg}.
As we will review in Appendix B,
the partition function
then becomes a kind of matrix model.
The factorization of these partition functions of 3d abelian theories
was found by Pasquetti in \cite{Pasquetti:2011fj}.
\begin{align}
Z^{U(1)}
=\sum_i Z_{\textrm{vort}}^{(i)}(q,x)\,\widetilde{Z}_{\textrm{anti-vort}}^{(i)}(\tilde{q},\tilde{x})
=\parallel Z_{\textrm{vort}}\parallel^2_{\textrm{S}},
\end{align}
where the operation $\widetilde{\cdots}$ 
in the sum
acts for instance as $q=e^{\hbar}\to\tilde{q}=e^{-1/\hbar}$.
So this pairing involves the S-duality transformation,
and we call it the $S$-pairing.
The geometric meaning of the S-transformation will be clear in this section.

The supersymmetric index is also important quantity
to catch a part of quantum dynamics of theory.
The 3d superconformal index,
which is defined for a 3d SCFT,
is the following trace taken over the Hilbert space of
the theory on $\mathbb{R}\times S^2$:
\begin{align}
I(q,z)
=\textrm{Tr}
(-1)^Fe^{-\beta\{\mathcal{Q},\mathcal{S}\}}
q^{\epsilon+j}
\prod_i
z_i^{F_i}
.
\end{align}
Here $F_i$ is a Cartan generator of the flavor symmetry.
The bosonic parti of the 3d $\mathcal{N}=2$ superconformal group 
is $SO(3,2)\times SO(2)$,
and the quantum numbers under the Cartan generators
of its compact subgroup $SO(2)_j\times SO(3)_\epsilon\times SO(2)_R$
label the states of the 3d theory.
Then the above superconformal index counts the BPS states for $\mathcal{Q}$ and $\mathcal{Q}^\dag=\mathcal{S}$:
\begin{align}
\{\mathcal{Q},\mathcal{S}\}
=-j+\epsilon-R\equiv0,
\end{align}
so this index does not depend on $\beta$,
and we can take the limit $\beta\to\infty$ to evaluate it.

On the one hand we can write down the index as 
a twisted partition function of a 3d theory 
defined on the curved space-time
$S^1\times S^2$,
\begin{align}
I(q,z)
=\int_{\textrm{twisted b.c.}} \mathcal{D}\Psi
e^{-S[{S^1\times S^2}]-t\{Q,V\}}
.
\end{align}
The index was also calculated by using the localization formula \cite{Kim:2009wb,Imamura:2011su},
and then the path integral reduces to an ordinary integral over the Cartan of the gauge group
via supersymmetric localization \cite{Iqbal:2012xm} .
Factorization for the resulting expression of the index was predicted in \cite{Dimofte:2011py}
\begin{align}
I^{U(N)}=
\sum_i
Z_{\textrm{vort}}^{(i)}(q,z)\cdot \overline{Z}_{\textrm{anti-vort}}^{(i)}(\bar{q},\bar{z})
=\parallel Z_{\textrm{vort}}\parallel^2_{{id}}.
\end{align}
In \cite{Beem:2012mb,Hwang:2012jh} this conjecture is studied in more detail.

For 3d abelian gauge theories,
it is observed that this building block $Z_{\textrm{vort}}^{(i)}$ of the index is identical with
that of the factorized partition function in \cite{Pasquetti:2011fj}.
The difference is the meaning of the conjugated variables $\bar{q}$ and $\bar{z}$.
The conjugation here is merely the inversion $\bar{q}=q^{-1}$,
and we call this paring the $id$-pairing.

\begin{figure}[tbp]
 \begin{center}
  \includegraphics[width=120mm,clip]{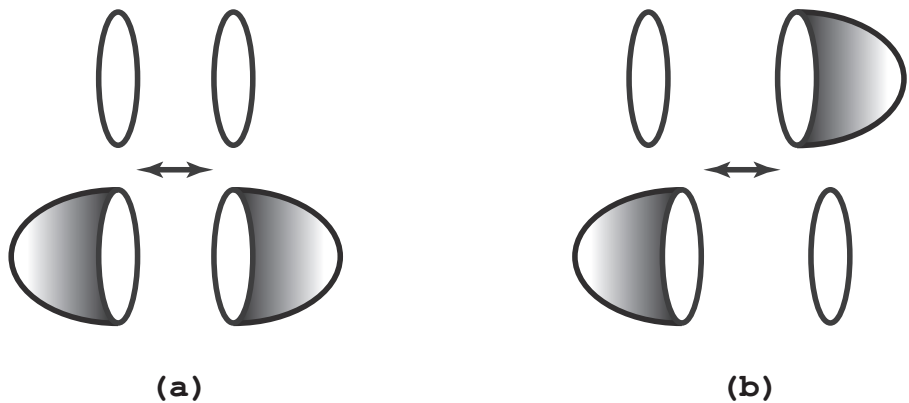}
 \end{center}
 \caption{(a) The $id$-gluing of two $\bar{T}^2$'s: the trivial decomposition of $S^1\times S^2$.
(b) The $S$-gluing of two $\bar{T}^2$'s: the Heegaard decomposition of three-sphere $S^3$
through the S element of the mapping class group $SL(2, \mathbb{Z})$.}
 \label{fig;dec}
\end{figure}
In \cite{Beem:2012mb}, it is proposed that
the above-mentioned factorization originates in geometry
on which the quantum field theory is defined.
As \emph{\textbf{Figure}} {\bi{\ref{fig;dec}}},
$S^1\times S^2$ and $S^3$ are $id$- and $S$-gluing of a pair of solid tori $S^1\times D^2=\bar{T}^2$.
Actually the squashed sphere $S^3_b$,
on which our discussion focuses,
is the $S$-gluing of two half geometries $S^1\times_q D^2$,
and this building block is the Melvin cigar \cite{Cecotti:2010fi,Beem:2012mb}
where $D^2$ fibers over $S^1$ with holonomy $q$.
In \cite{Beem:2012mb} the authors defined the holomorphic blocks
as the partition functions on this Melvin cigar $S^1\times_q D^2$.
This partition function is just the wave function 
for the Hilbert space
on the asymptotic $\mathbb{R}\times T^2$.
Here $\mathbb{R}$ is the infinite time direction,
and then a state evolves into a ground state as $\langle 0_q\vert i\rangle$.
In this way the wave function
depends on the choice of the supersymmetric vacuum $\vert i\rangle$
which specifies a state on the boundary $T^2$.
Then the gluing of two geometries through an element of $SL(2, \mathbb{Z})$
 implies the following form of the partition function
on the total geometry:
\begin{align}
Z
=\langle\,0_q\,\vert\,0_{\tilde{q}}\,\rangle
=\sum_i \langle\,0_q\,\vert i \,\rangle
\langle\,i\,\vert
\,0_{\tilde{q}}\,\rangle.
\end{align}
This is the geometric origin of the factorization which was conjectured in \cite{Beem:2012mb}.

As we mentioned, the factorization is actually observed in
abelian partition functions and non-abelian superconformal indexes.
So in order to verify this conjecture for a wide range of 3d theories,
we have to confirm the factorization phenomenon for partition functions
of non-abelian gauge theories.
In the following sections,
we show that non-abelian partition functions actually factorize
into the holomorphic blocks which are consistent with superconformal index \cite{Hwang:2012jh}
and have a nice 3d interpretation.


\section{Factorization of ellipsoid partition functions}

The following matrix model gives
the partition function of $\mcN=2$ $U(N)$ gauge theory
with $N_f$ fundamentals and $\bar{N}_f$ anti-fundamentals
on squashed three-sphere
\cite{Jafferis:2010un,Hama:2010av,Hama:2011ea,Imamura:2011wg} 
\begin{align}
Z
=
\frac{1}{N!}\int d^Nx\,\,
&e^{-i\pi k\sum x_\al^2+2\pi i\xi\sum x_\al}
\prod_{1\leq \al <\be\leq N}
4\sinh\pi b(x_\al-x_\be)\,\sinh\pi b^{-1}(x_\al-x_\be)
\nonumber\\
&\quad\times
\prod_{\al=1}^N\,\prod_{i=1}^{N_f} 
\,s_b(iQ/2-M_i+x_\al  )
\prod_{\al=1}^N\,\prod_{j=1}^{\bar{N}_f} 
\,s_b(iQ/2-\tilde{M}_j-x_\al  ),
\end{align}
where $k$ is the Chern-Simons coupling,
$\xi$ is the FI-parameter for the $U(1)$ factor,
and $b$ is the squashing parameter for the three-sphere.
We give masses for fundamental
and anti-fundamental chiral field.
When $N_f=\bar{N}_f$,
this theory is the mass deformation of $\mathcal{N}=3$ SQCD
and a pair of a fundamental and an anti-fundamental chirals forms a hypermultiplet
of the $\mathcal{N}=3$ theory.
The deformation of $M=\tilde{M}$ type
is the vector mass for the original hypermultiplet,
and $M=-\tilde{M}$ type is the axial mass.
In the following, we turn on these mass deformations.

\subsection{$\mathcal{N}=2$ $U(N)$ vector-like theory}

\begin{figure}[tbp]
 \begin{center}
  \includegraphics[width=50mm,clip]{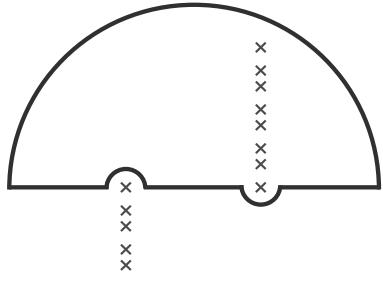}
 \end{center}
 \caption{The integral contour for the partition function. The crosses denote
 the poles of the integrant of the partition function.}
 \label{fig;cont}
\end{figure}

We start with studying $U(N)$ non-chiral gauge theory
whose matter content consists of $\mathcal{N}=2$ mass deformation of  $\mcN=3$ hypermultiplets.
The localized partition function is
\begin{align}
Z
=
\frac{1}{N!}\int d^Nx\,\,
e^{-i\pi k\sum x_\al^2+2\pi i\xi\sum x_\al}
\prod_{1\leq \al <\be\leq N}
&4\sinh\pi b(x_\al-x_\be)\,\sinh\pi b^{-1}(x_\al-x_\be)
\nonumber\\
&\times
\prod_{\al=1}^N\,\prod_{i=1}^{N_f} 
\,\frac{s_b(x_\al+m_i+\mu_i/2+ iQ/2 )}{s_b(x_\al+m_i-\mu_i/2- iQ/2  )}.
\end{align}
For $k=0$ we can enclose the integral contour in the upper half-plane as 
\emph{\textbf{Figure}} {\bi\ref{fig;cont}},
and in \cite{Pasquetti:2011fj} the author employed it for generic Chern-Simons coupling.
In this paper we follow the argument there.

To evaluate the matrix model for non-abelian theory $N\geq2$,
the best strategy is to translate this ``many-body" problem into that of abelian theory,
or a collection of non-interacting one-body systems.
This idea of ``abelianization" plays a role in solving many problems in for instance
\cite{Benvenuti:2011ga,Marino:2011eh},
and the Cauchy formula is the key in these articles.
The Cauchy formula implies
\begin{align}
\label{Cauh}
&\prod_{1\leq \al<\be\leq N}2\sinh(x_\al-x_\be)\nonumber\\
&=\frac{1}{\prod_{1\leq \al<\be\leq N}2\sinh(\chi_\al-\chi_\be)}
\sum_{\sigma\in S^N}(-1)^\sigma
\prod_{\al}\prod_{\be\neq\sigma(\al)}
2\cosh(x_\al-\chi_\be),
\end{align}
for auxiliary variables $\chi_\al\not\equiv\chi_\be$ (mod$\,\pi i$).
So with this formula,
we can resolve the``sinh-interaction" between ``particles" $x_\al$
into a collection of one-particle systems in a background $\chi_\al$.
We therefore use this formula as a separation of variables of our problem.

Substituting the formula (\ref{Cauh}) into the partition function,
we obtain the following expression:
\begin{align}
&Z
=\label{NCPF}
\frac{1}{N!
\prod_{1\leq \al<\be\leq N}4\sinh\pi b(\chi_\al-\chi_\be)
\sinh\pi b^{-1}(\chi'_\al-\chi'_\be)
}
\sum_{\sigma,\rho\in S^N}(-1)^{\sigma+\rho}
\prod_\al Z[\sigma(\al),\rho(\al)],
\\
&Z[\sigma(\al),\rho(\al)]=\oint dx\,\,
e^{-i\pi k x^2+2\pi i\xi x}
\prod_{\be\neq\sigma(\al)}2\cosh\pi b(x-\chi_\be)
\prod_{\be\neq\rho(\al)}2\cosh\pi b^{-1}(x-\chi'_\be)\nonumber\\
&\qquad\qquad\qquad\qquad\qquad\qquad\qquad\,\,
\label{pNCPF}
\times\prod_{i=1}^{N_f} 
\,\frac{s_b(x_\al+m_i+\mu_i/2+ iQ/2 )}{s_b(x_\al+m_i-\mu_i/2- iQ/2  )}.
\end{align}
Therefore the evaluation of the abelian integral $Z[\sigma(\al),\rho(\al)]$
immediately implies an explicit formula for the non-abelian partition function.
This integral is essentially equal to that was computed in \cite{Pasquetti:2011fj},
and we provide an explicit computation in Appendix C.
By computing this integral exactly,
we find the following factorized form of the partition function of the 3d non-chiral theory:
\begin{align}
Z=\frac{1}{N!}\sum_{i_1=1}^{N_f}\cdots\sum_{i_N=1}^{N_f}\,
Z_{\textrm{cl}}^{\{i_\al\}}\,
Z_{\textrm{pert}}^{\{i_\al\}}\,
Z_{V}^{\{i_\al\}}\,
\widetilde{Z}_{V}^{\{i_\al\}}.
\end{align}
Here the summation is taken over the sequece of integers  $\{i_\al=1,\cdots,N_f\}$,
which labels the supersymmetric ground states of the theory on $S^1\times\mathbb{R}^2$.
The perturbative part is given by
\begin{align}
&Z_{\textrm{cl}}^{\{i_\al\}}
(m,\mu,\xi)
=\prod_{\al=1}^Ne^{-i\pi k (m_{i_\al}+\mu_{i_\al}/2)^2-2\pi i\xi (m_{i_\al}+\mu_{i_\al})/2},\\
&Z_{\textrm{pert}}^{\{i_\al\}}
(m,\mu,b) 
=\prod_{1\leq\al<\be\leq N}
4\sinh(\pi bD_{i_\al i_\be})\,
4\sinh(\pi b^{-1}D_{i_\al i_\be})
\prod_{\al=1}^N\frac{\prod_{j\neq i_\al}^{N_f} s_b(D_{ji_\al}+ iQ/2 )}
{\prod_{j=1}^{N_f} s_b(C_{ji_\al}- iQ/2 )},
\end{align}
where
\begin{align}
D_{ji}=m_j-m_i+\mu_j/2-\mu_i/2,
\quad C_{ji}=m_j-m_i-\mu_j/2-\mu_i/2.
\end{align}
The remaining parts,
which are the holomorphic blocks, take the form
\begin{align}
&Z_{V}^{\{i_\al\}}
=
\sum_{m_1,\cdots,m_N=0}^\infty 
\prod_{\al=1}^N
\left((-1)^Ne^{\pi b\sum\mu_j}q^{N_f/2}z_\al\right)^{m_\al}q^{-km_\al^2/2}
\nonumber\\
&\times
\prod_{\al=1}^{N}\prod_{l=1}^{m_\al} \,\frac{
\prod_{j=1 }^{N_f}2\sinh \pi b (C_{j i_\al}+i(l-1)b)
}{
\prod_{\be=1}^N 2\sinh\pi b(D_{i_\al i_\be}+i(l-1-m_\al) b)
\prod_{j=1, \not\in\{i_\al\}}^{N_f}2\sinh \pi b (D_{j i_\al}+ilb)
},\\
&\widetilde{Z}_{V}^{\{i_\al\}}
=
\sum_{n_1,\cdots,n_N=0}^\infty 
\prod_{\al=1}^N
\left((-1)^Ne^{\pi b^{-1}\sum\mu_j}\tilde{q}^{N_f/2}\tilde{z}_\al\right)^{n_\al}\tilde{q}^{-kn_\al^2/2}
\nonumber\\
&\times
\prod_{\al=1}^{N}\prod_{\ell=1}^{n_\al} \,\frac{
\prod_{j=1 }^{N_f}2\sinh \pi b^{-1} (C_{j i_\al}+i(\ell-1)b^{-1})
}{
\prod_{\be=1}^N 2\sinh\pi b^{-1}(D_{i_\al i_\be}+i(\ell-1-n_\al) b^{-1})
\prod_{j=1, \not\in\{i_\al\}}^{N_f}2\sinh \pi b^{-1} (D_{j i_\al}+i\ell b^{-1})
}.
\end{align}
These blocks are precisely equal to the 3d (K-theoretic) uplift
of the vortex and anti-vortex partition functions \cite{Shadchin:2006yz}
for $U(N)$ gauge theory with $N_f$ antifundamental and $N_f-N$ fundamental
chiral multiplets.
The Coulomb branch and the mass parameters 
for the vortex theory are
\begin{align}
&\label{CBP}a_\al=m_{i_\al}+\frac{\mu_{i_\al}}{2},\quad \al=1,\cdots,N\\
&\label{MoF}M_j=m_j-\frac{\mu_{j}}{2},\quad j=1,\cdots ,N_f\\
&\label{MoAF}\bar{M}_{{j}}=m_{{j}}+\frac{\mu_{{{j}}}}{2}+ib^\pm,
\quad{j}\not\in \{i_1,\cdots,i_N\}.
\end{align}
See Appendix C for detailed computation and discussion.

The classical and 1-loop part $Z_{\textrm{cl}}^{\{i_\al\}}
Z_{\textrm{pert}}^{\{i_\al\}}$ is basically a product of those of $U(1)$ theory,
and we can show that
it is precisely the purtabative part of the 
vortex/antivortex partition function \cite{Shadchin:2006yz}.
Actually, we have the factorization of the 1-loop contributions \cite{Pasquetti:2011fj}
\begin{align}
Z_{\textrm{pert}}^{\{i_\al\}}
=
\prod_\al
e^{i\pi\sum_j ((D_{ji_\al}+iQ/2)^2-(C_{ji_\al}-iQ/2)^2)/2}\times
Z_{\textrm{1-loop}}^{\{i_\al\}}\,\widetilde{Z}_{\textrm{1-loop}}^{\{i_\al\}},
\end{align}
where
\begin{align}
&Z_{\textrm{1-loop}}^{\{i_\al\}}=
\prod_{1\leq\al<\be\leq N}
4\sinh(\pi bD_{i_\al i_\be})\,
\prod_\al\prod_{\ell=1}^\infty
\frac{\prod_{j\neq i_\al}(1-q^\ell e^{-2\pi bD_{ji_\al}})}
{\prod_{j}(1-q^{\ell-1} e^{-2\pi bC_{ji_\al}})},\\
&\widetilde{Z}_{\textrm{1-loop}}^{\{i_\al\}}
=
\prod_{1\leq\al<\be\leq N}
4\sinh(\pi b^{-1}D_{i_\al i_\be})\,
\prod_\al\prod_{\ell=1}^\infty
\frac{\prod_{j\neq i_\al}(1-\tilde{q}^\ell e^{-2\pi b^{-1}D_{ji_\al}})}
{\prod_{j}(1-\tilde{q}^{\ell-1} e^{-2\pi b^{-1}C_{ji_\al}})}.
\end{align}
The prefactor $\prod_\al e^{\pi i\sum (D+iQ/2)^2-(C-iQ/2)^2)}$
can be absorbed into the classical part,
up to irrelevant overall coefficient $e^{-i\pi N\sum m_j\mu_j}$,
by changing the FI parameters \cite{Pasquetti:2011fj}
\begin{align}
\xi\,\,\to\,\,\xi_{\textrm{eff}}=\xi+\frac{1}{2}\sum_j(\mu_j+iQ)
.
\end{align}
We thus obtain the factorization of the non-chiral $U(N)$ theory
as a natural extension of that of $U(1)$ theory: 
\begin{align}
Z=
\frac{1}{N!}\,
\sum_{i_1=1}^{N_f}\cdots\sum_{i_N=1}^{N_f}\,
Z_{\textrm{cl}}^{\{i_\al\}}(\xi_{\textrm{eff}})\,
\bigg(
Z_{\textrm{1-loop}}^{\{i_\al\}}Z_{V}^{\{i_\al\}}
\bigg)\bigg(
\widetilde{Z}_{\textrm{1-loop}}^{\{i_\al\}}
\widetilde{Z}_{V}^{\{i_\al\}}
\bigg).
\end{align}
As we had expected, the holomorphic block $Z_{\textrm{1-loop}}^{\{i_\al\}}Z_{V}^{\{i_\al\}}$
 coming from this factorization coincides with
that of the superconformal index of the same gauge theory \cite{Hwang:2012jh}.
We can therefore conclude that the single holomorphic block
leads to not only the partition function but also the superconformal index 
of the vector-like gauge theory.
As we will see in below, this fact holds for chiral theories.

In the next section,
we will see that this non-abelian holomorphic block $Z_{\textrm{1-loop}}^{\{i_\al\}}Z_{V}^{\{i_\al\}}$
we derived here
can be reformulated into an open topological string partition function
in the presence of $N$ A-branes on strip geometry.

\subsection{$\mathcal{N}=2$ $U(N)$ chiral theory}

We move on to studying chiral gauge theory.
In this section we deal with $\mcN=2$ $U(N)$ gauge theory
with $2N_f$ fundamental chiral multiplets.
The partition function is given by the following matrix model:
\begin{align}
Z
=
\frac{1}{N!}\int d^Nx\,\,
e^{-i\pi k\sum x_\al^2+2\pi i\xi\sum x_\al}
\prod_{1\leq \al <\be\leq N}
&4\sinh\pi b(x_\al-x_\be)\,\sinh\pi b^{-1}(x_\al-x_\be)
\nonumber\\
&\times
\prod_{\al=1}^N\,\prod_{i=1}^{2N_f} 
\,{s_b(x_\al+\mu_i/2+ iQ/2 )}.
\end{align}
Here we turn on the axial masses $\mu_i$ to the chiral multiplets 
by turning on the scalar VEVs
for the background vector multiplets
of weakly-gauged $U(1)$ symmetry.

This partition function also takes the factorized form
\begin{align}
Z=\frac{1}{N!}\,\sum_{i_1=1}^{2N_f}\cdots\sum_{i_N=1}^{2N_f}\,
Z_{\textrm{cl}}^{\{i_\al\}}\,
Z_{\textrm{pert}}^{\{i_\al\}}\,
Z_{V}^{\{i_\al\}}\,
\widetilde{Z}_{V}^{\{i_\al\}}.
\end{align}
See Appendix C for detailed computation.
The perturbative part is given by
\begin{align}
&Z_{\textrm{cl}}^{\{i_\al\}}
(m,\mu,\xi)
=\prod_{\al=1}^Ne^{-i\pi k (\mu_{i_\al}/2)^2-2\pi i\xi \mu_{i_\al}/2},\\
&Z_{\textrm{pert}}^{\{i_\al\}}
(m,\mu,b) 
=
\prod_{\al<\be}4\sinh\pi b(E_{i_\al i_\be})
\sinh\pi b^{-1}(E_{i_\al i_\be})\nonumber\\
&\qquad\qquad\quad\quad\quad
\times\prod_{\al=1}^N\prod_{j\neq i_\al}^{2N_f} s_b(E_{ji_\al }+ iQ/2 )
\propto Z_{\textrm{1-loop}}^{\{i_\al\}}\,\widetilde{Z}_{\textrm{1-loop}}^{\{i_\al\}},
\end{align}
where $E_{ji}=\mu_{j}/2-\mu_{i}/2$.
The proportional coefficient in the last line
can be absorbed into the classical part by the change of the couplings
\begin{align}
k\to k_{\textrm{eff}}=k+N_f,\quad
\xi\to \xi_{\textrm{eff}}=\xi+\frac{iQN_f}{2}.
\end{align}
The full partition function then take the following factorized form
\begin{align}
Z=\frac{1}{N!}\,\sum_{i_1=1}^{N_f}\cdots\sum_{i_N=1}^{N_f}\,
Z_{\textrm{cl}}^{\{i_\al\}}(k_{\textrm{eff}},\xi_{\textrm{eff}})\,
\left(
{Z}_{\textrm{1-loop}}^{\{i_\al\}}
Z_{V}^{\{i_\al\}}\right)\,
\left(
\widetilde{Z}_{\textrm{1-loop}}^{\{i_\al\}}
\widetilde{Z}_{V}^{\{i_\al\}}\right).
\end{align}
The holomorphic block for this chiral theory is
\begin{align}
&Z_{V}^{\{i_\al\}}
=
\sum_{m_1,\cdots,m_N=0}^\infty 
\prod_{\al=1}^N
\left((-1)^Nq^{N_f}z_\al\right)^{m_\al}q^{-km_\al^2/2}
\nonumber\\
&\times
\prod_{\al=1}^{N}\prod_{l=1}^{m_\al} \,
\frac{
1}{
\prod_{\be=1}^N 2\sinh\pi b(E_{i_\al i_\be}-ilb)
\prod_{j=1, \not\in\{i_\al\}}^{2N_f}2\sinh \pi b (E_{j i_\al}+ilb)
}.
\end{align}
The anti-vortex part $\widetilde{Z}_V^{\{i\}}$
is given by the replacement 
$b\to b^{-1}$, $q\to \tilde{q}$ and $z_\al\to \tilde{z}_\al$.
This is precisely equal to the vortex partition function
for $U(N)$ theory with $2N_f-N$ fundamental chital multiplets.
This is the non-abelian generalization of the holomorphic block
for $N=1$ theory \cite{Pasquetti:2011fj}.

\section{Vortex partition functions and topological strings}

\begin{figure}[tbp]
 \begin{center}
  \includegraphics[width=110mm,clip]{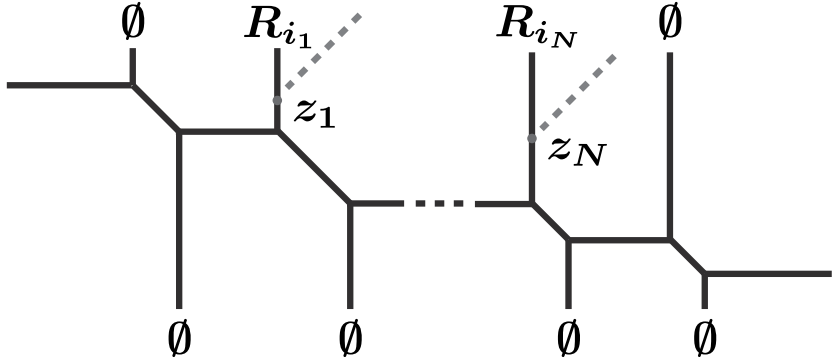}
 \end{center}
 \caption{This strip geometry is half of the toric geometry which leads to
 4d $\mcN=2$ $U(N_f)$ gauge theory with $2N_f$ flavors through the geometric engineering.
 We set $R_j=\emptyset$ for $j\not\in \{i_\al\}$.}
 \label{fig;str}
\end{figure}

In this section,
we provide an interpretation of 
non-abelian vortex partition functions
in terms of open topological string theory.
In \cite{Pestun:2007rz},
it was shown that the holomorphic blocks for
$U(1)$ gauge theories are given by
partition functions of
open topological strings
in the presence of single A-brane.
By generalizing this argument,
we demonstrate that
the open topological string partition functions
with multiple A-branes
give the holomorphic blocks for
non-abelian gauge theories.

The topological string partition function \cite{Aganagic:2003db}
of the strip geometry \emph{\textbf{Figure}} {\bi\ref{fig;str}}
which gives the Nekrasov partition function of
4d $\mcN=2$ gauge theory with $N_F=2N_C$ flavors is \cite{Iqbal:2004ne}
\begin{align}
\frac{\mathcal{K}_{\quad\emptyset\emptyset\cdots \emptyset}^{R_1R_2\cdots R_N}}
{\mathcal{K}_{\emptyset\emptyset\cdots \emptyset}^{\emptyset\emptyset\cdots \emptyset}}
=\prod_{i=1}^NS_{R_i}(q^{\rho})\prod_{\l=1}^\infty
\frac{\prod_{i\leq j}(1-q^lQ_{a_ib_j})^{C_l(R_i,\emptyset)}
\prod_{j< i}(1-q^lQ_{b_ja_i})^{C_l(\emptyset, R_i^T)}
}
{\prod_{i<i'}
(1-q^lQ_{a_ia_{i'}})^{C_l(R_i ,R_{i'}^T)}
},
\end{align}
where
\begin{align}
\sum_lC_l(Y,R)q^l
=q^{-1}(q-1)^2f_Y(q)f_R(q)+f_Y(q)+f_R(q),\quad
f_Y(q)=\sum_{(i,j)\in Y}q^{j-i}.
\end{align}
In the following we show that
the open string partition function on this geometry
gives the holomorphic block
for $U(N)$ non-chiral theory in the previous section.
Notice that this geometry is the same as that of $U(1)$ case \cite{Pestun:2007rz}.
The difference is the number of A-branes we insert in the geometry,
and we now consider $N$ branes for non-abelian gauge theory.
Let us consider A-brane insertion at $i_\al$-th legs of the strip geometry as
\emph{\textbf{Figure}} {\bi\ref{fig;str}}.
Since the world-sheet instanton on single A-brane is labeled by the Young diagrams $1^m$,
the following assignment of representations leads to the open string partition function
with instanton mode ${m_\al}$:
\begin{align}
R_{i_\al}=1^{m_\al}\,\,\textrm{for }i_{\al=1,2,\cdots, N},\quad
\textrm{otherwise}\quad R_j=\emptyset.
\end{align}
Let $t_\al$ be the open string moduli on the $i_\al$-th brane.
The open topological string partition function $Z=\sum_{\{i_\al\}}e^{\sum t_\al m_\al}Z^{\{m_\al\}}$
is given by the following partition function of strip geometry
with non-trivial representations $[1^{m_\al}]$:
\begin{align}
Z^{\{m_\al\}}&\equiv
\frac{
\mathcal{K}_{\quad\,\,\,\emptyset\emptyset\cdots \emptyset}
^{\emptyset \cdots1^{m_1}\cdots 1^{m_N}\cdots\emptyset}}
{\mathcal{K}_{\emptyset\emptyset\cdots \emptyset}^{\emptyset\emptyset\cdots \emptyset}}\nonumber\\
&=
\prod_{\al=1}^N
\prod_{l=1}^{m_\al}\frac{1}{1-q^l}
\prod_{\l=1}^{m_\al}
\frac{
\prod_{i_\al\leq j}(1-q^{l-1}Q_{a_{i_\al}b_j})
\prod_{j< i_\al}(1-q^{-l+1}Q_{b_ja_{i_\al}})
}
{
\prod_{i_\al <i_\be}
(1-q^{-l}Q_{a_{i_\al} a_{i_\be}})
\prod_{i_\al >i_\be}
(1-q^{l}Q_{a_{i_\be} a_{i_\al}})
}\nonumber\\
&\quad
\times
\frac{
1
}
{
\prod_{j>i_\al  ,\not\in\{i_\al\}}
(1-q^{1-l}Q_{a_{i_\al} b_j})
\prod_{j<i_\al,\not\in\{i_\al\}}
(1-q^{l-1}Q_{b_j a_{i_\al}})
}\nonumber\\
&=\prod_{\al=1}^N
\prod_{l=1}^{m_\al}\frac{1}{1-q^l}
\prod_{\l=1}^{m_\al}
\frac{
\prod_{ j}(1-q^{l-1}Q_{a_{i_\al}b_j})
}
{
\prod_{i_\al ,i_\be}
(1-q^{-l}Q_{a_{i_\al} a_{i_\be}})
\prod_{j\not\in\{i_\al\}}
(1-q^{1-l}Q_{a_{i_\al} b_j})
}.
\end{align}
Under the identification between parameters,
\begin{align}
Q_{a_{i_\al} b_j}=e^{-2\pi bC_{ji_\al}},\quad
Q_{b_j a_{i_\al}}=qe^{-2\pi bD_{ji_\al}},\quad
Q_{a_{i_\al} a_{i_\be}}=e^{-2\pi bD_{i_\al i_\be}},
\end{align}
this open string partition function
is precisely the holomorphic block for
3d $U(N)$ vector-like theory
up to an overall monomial which is not
relevant for our discussion.
In other words,
the open topological string partition function
on the strip geometry
gives
the K-theoretic uplift of the vortex partition function
for $U(N)$ theory with $N_f$ antifundamentals and
$N_f-N$ fundamentals.
This is a generalization of
the relation between the vortex partition function
and topological strings found in \cite{Bonelli:2011fq} 
where the authors studied the special case $N=N_f$.

\begin{figure}[tbp]
 \begin{center}
  \includegraphics[width=110mm,clip]{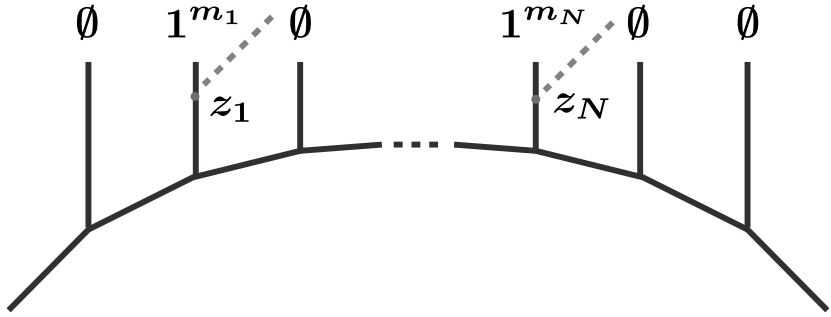}
 \end{center}
 \caption{This strip geometry is half of the toric geometry which leads to
 4d $\mcN=2$ $U(2N_f)$ pure Yang-Mills theory through the geometric engineering.}
 \label{fig;str}
\end{figure}
It is straightforward to generalize this computation for $U(N)$ 
chiral theory.
The relevant geometry for this case is the half $SU(2N_f)$ geometry,
and this geometry is the same as that of $U(1)$ case \cite{Pestun:2007rz} again.
We skip the detailed computation since it 
is merely a slight modification of the above case,
but it is easy to see that
the corresponding partition function gives
the holomorphic block for the $U(N)$ gauge theory
with $2N_f$ fundamentals.


\section{Discussion}
In this article,
we computed exactly
the supersymmetric partition functions of $\mcN=2$ $U(N)$ gauge theories
on the squashed three-sphere.
We then found that the resulting expression shows the factorized structure,
and it leads to the expected holomorphic block
of the 3d theory.
In this way
we gave an explicit proof of the factorization conjecture 
for a range of non-abelian gauge theories.
The obtained holomorphic blocks are consistent with the 
computation of the superconformal index,
and we found that
the blocks can be recast  into open topological string partition functions
with $N$ A-branes.

The system of the vortex counting of a 2d theory 
coupled to a bulk 4d $\mcN=2$ gauge theory
describes the surface operator of the 4d theory
\cite{Bonelli:2011fq,Alday:2009fs,Dimofte:2010tz,Kozcaz:2010af,Taki:2010bj,Awata:2010bz,Bonelli:2011wx,Chen:2013dda}.
Since the 3d uplift of the vortex theory gives the holomorphic block,
it is very interesting to investigate the 5d uplift of the surface operator, which is
3d gauge theory coupled to 5d gauge theory.
In this way we can synthesize the factorizations of
4d theory (\ref{Pestun}) and 3d theory (\ref{Pasquetti}),
and then this interplay of them should lead to 
 new phenomenon in 5d.
From the perspective of the AGT correspondence,
$q$-Toda theory will play a role in this direction
\cite{Nieri:2013yra,Kozcaz:2010af,Bao:2011rc}.
It should be interesting to study this issue further.

\section*{Acknowledgements}
The author would like to thank N.Hama for useful discussions.
The work is supported by Special Postdoctoral Researchers Program at RIKEN. .


\appendix
\def\thesection{Appendix \Alph{section}}
\section{Double-sine function}
\renewcommand{\theequation}{A.\arabic{equation}}
\setcounter{equation}{0}

 In \cite{Jafferis:2010un},
 Jafferis found that the following $\ell$-function plays a role in the localization computation
 and the F-extremization of three-dimensional theories:
\begin{align}
\ell(z)=
-z\log (1-e^{2\pi iz})+\frac{i}{2}\bigg(\pi z^2+\frac{1}{\pi}\textrm{Li}_2 (e^{2\pi iz})\bigg)
-\frac{i\pi}{12},
\end{align}
whose defining property is
\begin{align}
\frac{d\ell}{dz}=-\pi z\cot \pi z,\quad \ell(0)=0.
\end{align}
Let us consider the function $s(x)=e^{\ell(-ix)}$.
This function satisfies many nice properties,
and we can show that this is a specialization of the double-sine function $s_b$:
\begin{align}
s(x)=s_{b=1}(x).
\end{align}
The double-sine function is defined as a natural extension of the sine function
through the product expression \cite{Shintani}
\begin{align}
s_b(x)=\prod_{m,n=0}^\infty \frac{mb+n/b+Q/2-ix}{mb+n/b+Q/2+ix}.
\end{align} 
We can recast this definition into 
the language of Barnes gamma function $\Gamma_b(x)=\Gamma_2(x\vert b,b^{-1})$ \cite{Barnes}
\begin{align}
s_b(x)&=\frac{\Gamma_b(Q/2+ix)}{\Gamma_b(Q/2-ix)}.
\end{align}
From the definition, we can easily see the inversion relation
$s_b(x)={1}/{s_b(-x)}$.

The double-sine function  is meromorphic,
and it satisfies the following properties
\begin{align}
\frac{s_b(x+iQ/2+imb+in/b)}{s_b(x+iQ/2)}
&=\frac{(-1)^{mn}}{
\prod_{k=1}^{m}2i\sinh \pi b(x+ikb)
\prod_{\ell=1}^{n}2i\sinh \pi b(x+i\ell/b)}
\nonumber\\
&=\label{F1}
\frac{(-1)^{mn}(-i)^{m+n}q^{m(m+1)/4}\bar{q}^{n(n+1)/4} e^{-\pi bmx}e^{-\pi nx/b}}{
\prod_{k=1}^{m}(1-q^ke^{-2\pi bx})
\prod_{\ell=1}^{n}(1-\tilde{q}^\ell e^{-2\pi x/b})},
\\
\frac{s_b(x-iQ/2+imb+in/b)}{s_b(x-iQ/2)}
&=\frac{(-1)^{mn}}{
\prod_{k=1}^{m}2i\sinh \pi b(x-iQ+ikb)
\prod_{\ell=1}^{n}2i\sinh \pi b(x-iQ+i\ell/b)}
\nonumber\\
&=\label{F2}
\frac{(-1)^{mn}(-i)^{m+n}q^{m(m+1)/4}\bar{q}^{n(n+1)/4} e^{-\pi bm(x-iQ)}e^{-\pi n(x-iQ)/b}}{
\prod_{k=1}^{m}(1-q^{k-1}e^{-2\pi bx})
\prod_{\ell=1}^{n}(1-\tilde{q}^{\ell-1} e^{-2\pi x/b})},
\end{align}
where we introduce
\begin{align}
q=q_1^{-2}=e^{-2\pi ib^2},\quad
\tilde{q}= q_2^{-2}=e^{-2\pi i/b^2}.
\end{align}
These formulas come from the following expression of this function:
\begin{align}
s_b(x)
=
\frac{ e^{-i\pi /2x^2}}{
\prod_{k=1}^{\infty}(1-q^{k-1/2}e^{-2\pi bx})
\prod_{\ell=1}^{\infty}(1-\tilde{q}^{\ell-1/2} e^{-2\pi x/b})}.
\end{align}
We also have the integral representation
\begin{align}
\log s_b(ix)
=
\int_0^\infty \frac{dt}{t}
\bigg( \frac{\sinh 2tx}{2\sinh bt\sinh t/b}-\frac{x}{t}\bigg).
\end{align}
The residue is $\textrm{Res}_{x=iQ/2}\, s_b(x) = 1/2\pi$.

\section{3d partition function}
\renewcommand{\theequation}{B.\arabic{equation}}
\setcounter{equation}{0}

The supersymmetric localization of gauge theories on $S^3$
enables us to compute
their partition functions
as the conventional matrix model over the gauge group \cite{Jafferis:2010un,Hama:2010av}:
\begin{align}
Z=\frac{1}{\vert \mathcal{W}\vert} \int \prod_H \,d\sigma \,
Z^{\textrm{cl}}(\sigma)
Z_{\textrm{1-loop}}^{\textrm{vector}}(\sigma)
Z_{\textrm{1-loop}}^{\textrm{chiral}}(\sigma).
\end{align}
Here the``dynamical" variable $\sigma=\sum\sigma_iH_i$ 
originates from the auxiliary scalar component
of the vector multiplet.
The localization reduces the path integral
onto the constant VEV of the scalars.
Then the saddle point approximation is exact,
and the one-loop computation provides the factor 
$Z_{\textrm{1-loop}}(\sigma)$

\begin{align}
Z_{\textrm{1-loop}}^{\textrm{vector}}(\sigma)
=\det{}_{\textrm{Ad}}\,
\frac{2\sinh \pi\sigma}{\pi\sigma}
=\prod_{\al\in \Delta_{+}}
\frac{(2\sinh \pi \al_i\sigma_i)^2}{(\pi \al_i\sigma_i)^2}.
\end{align}
The denominator cancels with the Vandermonde determinant when we replace the matrix integral
with the eigenvalue integral $\int \prod_Hd\sigma\to\int\prod_i d\sigma_i$.

The chiral multiplet in the representation $R$
with R-charge (conformal dimension) $q$
gives the contribution
\begin{align}
Z_{\textrm{1-loop}}^{\textrm{chiral}}(\sigma)
=\det{}_{{R}}\,\,
e^{\ell(1-q+i\sigma)}
=\prod_{\rho\in R}
s_{b=1}(i-iq-\rho_i\sigma_i).
\end{align}

These results are generalized to the theories on the squashed sphere $S^3_b$
\cite{Hama:2011ea,Imamura:2011wg}.
There are some realizations of squashed three-sphere $S^3_b$,
and each preserves different subgroup of the isometry of $S^3$.
In this article we adopt that of the last section of \cite{Hama:2011ea},
which is natural in our context.
The building blocks of the partition functions then receive
a slight modification by the squashing parameter $b$:
\begin{align}
Z_{\textrm{1-loop}}^{\textrm{vector}}(\sigma)
&
=\prod_{\al\in \Delta_{+}}
\frac{4\sinh \pi b\al_i\sigma_i\,\sinh \pi b^{-1}\al_i\sigma_i}{(\pi \al_i\sigma_i)^2},\\
Z_{\textrm{1-loop}}^{\textrm{chiral}}(\sigma)
&
=\prod_{\rho\in R}
s_{b}(iQ/2-iq-\rho_i\sigma_i).
\end{align}

\section{Details of computation}
\renewcommand{\theequation}{C.\arabic{equation}}
\setcounter{equation}{0}

\subsection{vector-like theory}

In this section we provide detailed computation of the partition function (\ref{NCPF})
of $\mcN=2$ non-chiral gauge theory.
For the purpose,
we start with computing (\ref{pNCPF}).
Recall that the simple poles and zeros of the double-sine function $s_b(x)$ are
\begin{align}
&\textrm{simple poles}\quad:\quad x=i(mb+n/b+Q/2),\\
&\textrm{zeros}\qquad\qquad\,:\quad x=-i(mb+n/b+Q/2),
\end{align}
for the non-negative integers $m,n=0,1,\cdots$.
The simple poles in the upper-half plane therefore 
come from the double-sine functions in the numerator of (\ref{pNCPF}).
Therefore 
by collecting the contribution from the pole $x_i^{mn}=-m_i-\mu_i/2+i(mb+n/b)$,
we obtain
\begin{align}
Z[\sigma(\al),\rho(\al)]&=
\sum_{i=1}^{N_f}\sum_{m,n=0}^\infty
e^{-i\pi k (x_i^{mn})^2+2\pi i\xi x_i^{mn}}\nonumber\\
&\times
\prod_{\be\neq\sigma(\al)}2\cosh\pi b(x_i^{mn}-\chi_\be)
\prod_{\be\neq\rho(\al)}2\cosh\pi b^{-1}(x_i^{mn}-\chi'_\be)
\nonumber\\
&\times
\frac{\prod_{j\neq i}^{N_f} s_b(D_{ji}^{mn}+ iQ/2 )}{\prod_{j=1}^{N_f} s_b(C_{ji}^{mn}- iQ/2 )}
\,R^{mn}(b),
\end{align}
where
\begin{align}
&D_{ji}^{mn}=m_j-m_i+\mu_j/2-\mu_i/2+i(mb+n/b)
=D_{ji}+i(mb+n/b),
\\
&C_{ji}^{mn}=m_j-m_i-\mu_j/2-\mu_i/2+i(mb+n/b)
=C_{ji}+i(mb+n/b),
\\
&R^{mn}(b)=2\pi \,\textrm{Res}_{x=0}s_b(x+i(mb+n/b)+iQ/2)\nonumber\\
&\qquad\quad\,=
\frac{(-1)^{mn}(-i)^{m+n}q^{m(m+1)/4}\bar{q}^{n(n+1)/4} }{
\prod_{k=1}^{m}(1-q^k)
\prod_{\ell=1}^{n}(1-\tilde{q}^\ell)}.
\end{align}
Using the formulas (\ref{F1}) and (\ref{F2}),
we can rewrite it into the following form
\begin{align}
&Z[\sigma(\al),\rho(\al)]\nonumber\\
&=
\sum_{i=1}^{N_f}\sum_{m,n=0}^\infty
e^{-i\pi k (m_i+\mu_i/2)^2-2\pi i\xi (m_i+\mu_i/2)}
(-1)^{(N-1)(m+n)}q^{-km^2/2}\tilde{q}^{-kn^2/2}
\nonumber\\
&\times
e^{-2\pi bkm(m_i+\mu_i/2)
-m\pi b\sum_j(\mu_j+iQ)-2m\pi \xi b}
e^{
-2\pi b^{-1}kn(m_i+\mu_i/2)
-n\pi/b\sum_j(\mu_j+iQ)-2n\pi \xi /b}\,
\nonumber\\
&\times
\frac{\prod_{j\neq i}^{N_f} s_b(D_{ji}+ iQ/2 )}{\prod_{j=1}^{N_f} s_b(C_{ji}- iQ/2 )}
\,\frac{
\prod_{j}
\prod_{k=1}^{m}(1-q^{k-1} e^{-2\pi bC_{ji} })
\prod_{\ell=1}^{n}(1-\tilde{q}^{\ell-1} e^{-2\pi b^{-1}C_{ji}})
}{
\prod_{j
}
\prod_{k=1}^{m}(1-q^k e^{-2\pi bD_{ji} })
\prod_{\ell=1}^{n}(1-\tilde{q}^\ell e^{-2\pi b^{-1}D_{ji}})
}
\nonumber\\
&\times
\prod_{\be\neq\sigma(\al)}2\cosh\pi b(m_i+\mu_i/2-imb+\chi_\be)
\prod_{\be\neq\rho(\al)}2\cosh\pi b^{-1}(m_i+\mu_i/2-in/b+\chi'_\be),
\end{align}
for integral Chern-Simons coupling $k\in\mathbb{Z}$.
Notice that the property $\cosh \pi b (x+in/b)=(-1)^{n}\cosh \pi b x$
prevents the partition function from mixing $m$ and $n$ sectors.
This technical mechanism enables us to factorize the partition function
into the vortex and antivortex blocks.

The abelianized partition function then takes the following factorized form
\begin{align}
&Z[\sigma(\al),\rho(\al)]\nonumber\\
&=
\sum_{i=1}^{N_f}
z_{\textrm{cl}}^{(i)}(m,\mu) \,
z_{\textrm{pert}}^{(i)}(m,\mu,b) \,
z_{{V}}^{(i)}(m,\mu,q;\sigma(\al),\chi)
 \,
\tilde{z}_{{V}}^{(i)}(m,\mu,\tilde{q};\rho(\al),\chi').
\end{align}
The mechanism of this factorization is basically identical to that of abelian theory.
Actually we find the following expressions
for the classical and one-loop part are the same as those of $U(1)$ theory:
\begin{align}
&z_{\textrm{cl}}^{(i)}(m,\mu)
=e^{-i\pi k (m_i+\mu_i/2)^2-2\pi i\xi (m_i+\mu_i/2)},\\
&z_{\textrm{pert}}^{(i)}(m,\mu,b) 
=\frac{\prod_{j\neq i}^{N_f} s_b(D_{ji}+ iQ/2 )}{\prod_{j=1}^{N_f} s_b(C_{ji}- iQ/2 )}.
\end{align}
The vortex/anti-vortex part is given by
\begin{align}
&z_{{V}}^{(i)}(m,\mu,q;\sigma(\al),\chi)\nonumber\\
&
=\sum_{m=0}^\infty z_i^m
(-1)^{(N-1)m}q^{-km^2/2}\,
\prod_{j}\,\frac{
\prod_{k=1}^{m}(1-q^{k-1} e^{-2\pi bC_{ji} })
}{
\prod_{k=1}^{m}(1-q^k e^{-2\pi bD_{ji} })
}
\nonumber\\
&\qquad\qquad\qquad\qquad\qquad\quad\times
\prod_{\be\neq\sigma(\al)}2\cosh\pi b(m_i+\mu_i/2-imb+\chi_\be),
\end{align}
\begin{align}
&\tilde{z}_{{V}}^{(i)}(m,\mu,\tilde{q};\rho(\al),\chi')\nonumber\\
&
=\sum_{n=0}^\infty \tilde{z}_i^n
(-1)^{(N-1)n}\tilde{q}^{-kn^2/2}\,
\prod_{j}\,\frac{
\prod_{\ell=1}^{n}(1-\tilde{q}^{\ell-1} e^{-2\pi b^{-1}C_{ji}})
}{
\prod_{\ell=1}^{n}(1-\tilde{q}^\ell e^{-2\pi b^{-1}D_{ji}})
}
\nonumber\\
&\qquad\qquad\qquad\qquad\qquad\quad\times
\prod_{\be\neq\rho(\al)}2\cosh\pi b^{-1}(m_i+\mu_i/2-in/b+\chi'_\be).
\end{align}
Here we introduced
\begin{align}
&z_i=e^{-2\pi bk(m_i+\mu_i/2)
-\pi b\sum_j(\mu_j+iQ)-2\pi \xi b}
=e^{-2\pi bk(m_i+\mu_i/2)
-2\pi \xi_{\textrm{eff}} b},\\
&\tilde{z}_i=e^{-2\pi b^{-1}k(m_i+\mu_i/2)
-\pi b^{-1}\sum_j(\mu_j+iQ)-2\pi \xi b^{-1}}
=e^{-2\pi b^{-1}k(m_i+\mu_i/2)
-2\pi \xi_{\textrm{eff}} b^{-1}}.
\end{align}

Since the full partition function consists of 
the abelianized partition functions (\ref{pNCPF}),
the full partition function takes the following factorized form
\begin{align}
Z
=\frac{1}{N!}\,\sum_{\{i_\al\}}\, Z_{\textrm{cl}}^{\{i_\al\}}\, Z_{\textrm{pert}}^{\{i_\al\}}\,
Z_{V}^{\{i_\al\}}\, \widetilde{Z}_{V}^{\{i_\al\}},
\end{align}
where the summation is taken over
the vortex sector which is labelled by the sequence of integers 
$\{\,i_1,\cdots,i_{N}\vert\, i_\al=1,\cdots,N_f\}$.
The classical and perturbative part of the partition functions
are basically the products of 
over all the abelian contributions
\begin{align}
&Z_{\textrm{cl}}^{\{i_\al\}}=
\prod_{\al=1}^N\,
z_{\textrm{cl}}^{(i_\al)},
\\
&Z_{\textrm{pert}}^{\{i_\al\}}=
\prod_{1\leq\al<\be\leq N}
4\sinh(\pi bD_{i_\al i_\be})\,
4\sinh(\pi b^{-1}D_{i_\al i_\be})\,\,
\prod_{\al=1}^N\,
z_{\textrm{pert}}^{(i_\al)}.
\end{align}
The origin of the perturbative contribution
 from non-abelian vector multiplet 
 \begin{align}
 \label{pertvec}\prod_{1\leq\al<\be\leq N}
4\sinh(\pi bD_{i_\al i_\be})\,
4\sinh(\pi b^{-1}D_{i_\al i_\be}),\end{align}
will be clear in the following discussion.
As we saw in section.3, 
we can factorize the perturbative part as follows
\begin{align}
Z_{\textrm{cl}}^{\{i_\al\}}(\xi)\,Z_{\textrm{pert}}^{\{i_\al\}}=
Z_{\textrm{cl}}^{\{i_\al\}}(\xi_{\textrm{eff}})\,
Z_{\textrm{1-loop}}^{\{i_\al\}}\,
\widetilde{Z}_{\textrm{1-loop}}^{\{i_\al\}}.
\end{align}
The vortexx/antivortex part is more complicated.
Since the Cauchy formula involves a summation over the permutation,
we have the following expression
\begin{align}
&Z_{V}^{\{i_\al\}}
=\frac{\sum_{\sigma}(-1)^\sigma
\prod_{\al=1}^N\,
z_{{V}}^{(i_\al)}(m,\mu,q;\sigma(\al),\chi)}
{\prod_{1\leq \al<\be\leq N}4\sinh\pi b(-\chi_\al+\chi_\be)\,
\sinh(\pi bD_{i_\al i_\be})
},\\
&\widetilde{Z}_{V}^{\{i_\al\}}
=\frac{\sum_{\rho}(-1)^\rho
\prod_{\al=1}^N\,
\tilde{z}_{{V}}^{(i_\al)}(m,\mu,\tilde{q};\rho(\al),\chi')}
{\prod_{1\leq \al<\be\leq N}4\sinh\pi b^{-1}(-\chi'_\al+\chi'_\be)\,
\sinh(\pi b^{-1}D_{i_\al i_\be})}.
\end{align}
From the construction,
the full partition function is independent of the choice of the
auxiliary parameters $\chi$ and $\chi'$.
Actually,
taking the summation over the permutations $\sigma$ by using the Cauchy formula again,
we get 
\begin{align}
&Z_{V}^{\{i_\al\}}
=
\sum_{m_1,\cdots,m_N=0}^\infty 
\frac{\prod_{\al=1}^Nz_\al^{m_\al}(-1)^{(N-1)m_\al}q^{-km_\al^2/2}}
{\prod_{1\leq \al<\be\leq N}2
\sinh(\pi bD_{i_\al i_\be})}
\,
\prod_{j,\al}\,\frac{
\prod_{k=1}^{m_\al}(1-q^{k-1} e^{-2\pi bC_{j i_\al} })
}{
\prod_{k=1}^{m_\al}(1-q^k e^{-2\pi bD_{j i_\al} })
}
\nonumber\\
&\times
\frac{1
}
{\prod_{1\leq \al<\be\leq N}2\sinh\pi b(-\chi_\al+\chi_\be)}
\sum_{\sigma}(-1)^\sigma
\prod_{\al=1}^N\,
\prod_{\be\neq\sigma(\al)}2\cosh\pi b(m_{i_\al}+\mu_{i_\al}/2-im_\al b+\chi_\be)\
\nonumber
\\
&=
\sum_{m_1,\cdots,m_N=0}^\infty 
\prod_{\al=1}^Nz_\al^{m_\al}(-1)^{(N-1)m_\al}q^{-km_\al^2/2}\,
\prod_{\al=1}^N\prod_{j}\,\frac{
\prod_{k=1}^{m_\al}(1-q^{k-1} e^{-2\pi bC_{j i_\al} })
}{
\prod_{k=1}^{m_\al}(1-q^k e^{-2\pi bD_{j i_\al} })
}\nonumber\\
&\qquad\qquad\qquad\qquad\qquad\qquad\qquad\qquad\quad\times 
\prod_{\al<\be}\frac{2\sinh\pi b(D_{i_\al i_\be}-im_\al b+im_\be b)}
{2\sinh(\pi bD_{i_\al i_\be})}.
\end{align}
We can show that
this vortex partition function is precisely the ``K-theoretic" uplift of the vortex partition function
for $U(N)$ gauge theory with $N_f$ antifundamentals and $N_f-N$ fundamentals
which was derived by Shadchin \cite{Shadchin:2006yz}.
Let us check this agreement.
With some algebra, the holomorphic block takes the form
\begin{align}
&Z_{V}^{\{i_\al\}}
=
\sum_{m_1,\cdots,m_N=0}^\infty 
\prod_{\al=1}^Nz_\al^{m_\al}(-1)^{(N-1)m_\al}q^{-km_\al^2/2+N_fm_\al/2}
e^{\pi bm_\al \sum_j\mu_j}
\nonumber\\
&\times
\prod_{\al<\be}\frac{2\sinh\pi b(D_{i_\al i_\be}-im_\al b+im_\be b)}
{2\sinh(\pi bD_{i_\al i_\be})}
\prod_{\al=1}^N\prod_{i_\al=1}^{N_f}\,\frac{
\prod_{l=1}^{m_\al} 
2\sinh \pi b (C_{j_\al i_\al}+i(l-1)b)
}{
\prod_{l=1}^{m_\al} 
2\sinh \pi b (D_{j_\al i_\al}+ilb)
}
\end{align}
As explained in Appendix.B of 
\cite{Hwang:2012jh}\footnote{Such computation is ubiquitous 
in the study of the Nekrasov partition functions.
See also \cite{Eguchi:2003sj} for instance.},
we can rewrite the partition function into the following form
\begin{align}
&Z_{V}^{\{i_\al\}}
=
\sum_{m_1,\cdots,m_N=0}^\infty 
\prod_{\al=1}^N
\left((-1)^Ne^{\pi b\sum\mu_j}q^{N_f/2}z_\al\right)^{m_\al}q^{-km_\al^2/2}
\nonumber\\
&\times
\prod_{\al=1}^{N}\prod_{l=1}^{m_\al} \,\frac{
\prod_{j=1 }^{N_f}2\sinh \pi b (C_{j i_\al}+i(l-1)b)
}{
\prod_{\be=1}^N 2\sinh\pi b(D_{i_\al i_\be}+i(l-1-m_\al) b)
\prod_{j=1, \not\in\{i_\al\}}^{N_f}2\sinh \pi b (D_{j i_\al}+ilb)
}.
\end{align}
Meanwhile, with the parametrization
(\ref{CBP}), (\ref{MoF}) and (\ref{MoAF}),
we can rewrite the arguments of the sinh factors
\begin{align}
C_{ji_\al}=M_j-a_\al,\quad
D_{{j}i_\al}=\bar{M}_{{j}}-a_\al+ib,\quad
D_{i_\al i_\be}=a_\al-a_\be.
\end{align}
Then, it is easy to see that this is precisely the vortex partition function 
of the above-mentioned $U(N)$ theory on $S^1\times\mathbb{R}^2$
whose chiral multiplets have masses $M_j$ and $\tilde{M}_j$.

\subsection{chiral theory}

Let us move on to the chiral gauge theory.
Our theory is $U(N)$ theory with $2N_F$ fundamental chiral multiplets,
and the axial masses satisfy $\sum_{i=j}^{2N_f}m_j=0$.

After using the Cauchy formula and 
performing the
integral using the residues at the simple poles
$x^{mn}_i=-\mu_i/2+i(mb+n/b)$,
the abelianized partition function is
\begin{align}
Z[\sigma(\al),\rho(\al)]&=
\sum_{i=1}^{N_f}\sum_{m,n=0}^\infty
e^{-i\pi k (x_i^{mn})^2+2\pi i\xi x_i^{mn}}\nonumber\\
&\times
\prod_{\be\neq\sigma(\al)}2\cosh\pi b(x_i^{mn}-\chi_\be)
\prod_{\be\neq\rho(\al)}2\cosh\pi b^{-1}(x_i^{mn}-\chi'_\be)
\nonumber\\
&\times
{\prod_{j\neq i}^{2N_f} s_b(E_{ji}^{mn}+ iQ/2 )}
\,R^{mn}(b),
\end{align}
where $E_{ji}=\mu_j/2-\mu_i/2$.
With the formulas (\ref{F1}) and (\ref{F2}),
we can rewrite it as
\begin{align}
&Z[\sigma(\al),\rho(\al)]\nonumber\\
&=
\sum_{i=1}^{N_f}\sum_{m,n=0}^\infty
e^{-i\pi k (\mu_i/2)^2-2\pi i\xi (\mu_i/2)}
(-1)^{(N-1)(m+n)}q^{-km^2/2}\tilde{q}^{-kn^2/2}
\nonumber\\
&\times
e^{-2\pi mb
\left(
\xi_{\textrm{eff}}
+\frac{\mu_i}{2}k_{\textrm{eff}}
-2\frac{iN_fQ}{2}
\right)}
e^{-2\pi nb^{-1}
\left(
\xi_{\textrm{eff}}
+\frac{\mu_i}{2}k_{\textrm{eff}}
-2\frac{iN_fQ}{2}
\right)}
\nonumber\\
&\times
\frac{\prod_{j\neq i}^{N_f} s_b(E_{ji}+ iQ/2 )}
{
\prod_{j
}
\prod_{k=1}^{m}(1-q^k e^{-2\pi bE_{ji} })
\prod_{\ell=1}^{n}(1-\tilde{q}^\ell e^{-2\pi b^{-1}E_{ji}})
}
\nonumber\\
&\times
\prod_{\be\neq\sigma(\al)}2\cosh\pi b(\mu_i/2-imb+\chi_\be)
\prod_{\be\neq\rho(\al)}2\cosh\pi b^{-1}(\mu_i/2-in/b+\chi'_\be)
\end{align}
for integral Chern-Simons coupling $k\in\mathbb{Z}$.
From this expression
we can obtain the following factorization as in the case of the vector-like theory
\begin{align}
Z
=\frac{1}{N!}\,\sum_{\{i_\al\}}\, Z_{\textrm{cl}}^{\{i_\al\}}\, Z_{\textrm{pert}}^{\{i_\al\}}\,
Z_{V}^{\{i_\al\}}\, \widetilde{Z}_{V}^{\{i_\al\}}.
\end{align}
The holomorphic block for this chiral theory is 
\begin{align}
&Z_{V}^{\{i_\al\}}
=
\sum_{m_1,\cdots,m_N=0}^\infty 
\prod_{\al=1}^{N}z_\al^{m_\al}(-1)^{(N-1)m_\al}q^{-km_\al^2/2}\,
\prod_{j=1}^{2N_f}
\prod_{\al=1}^N
\frac{
1
}{
\prod_{l=1}^{m_\al}(1-q^l e^{-2\pi b(\mu_{j}/2-\mu_{i_\al}/2) })
}\nonumber\\
&\qquad\qquad\qquad\qquad\qquad\qquad\times 
\prod_{\al<\be}\frac{2\sinh\pi b(\mu_{i_\al}/2-\mu_{i_\be}/2-im_\al b+im_\be b)}
{2\sinh\pi b(\mu_{i_\al}/2-\mu_{i_\be}/2)}.
\end{align}
where
\begin{align}
z_{\al}=
e^{
-2\pi b\left(k_{\textrm{eff}}\frac{\mu_{i_\al}}{2}+\xi_{\textrm{eff}}-\frac{iQN_f}{2}
\right)},\quad
\tilde{z}_{\al}
=e^{
-2\pi b^{-1}
\left(
k_{\textrm{eff}}
\frac{\mu_{i_\al}}{2}
+\xi_{\textrm{eff}}
-\frac{iQN_f}{2}
\right)
}.
\end{align}
Using the result in Appendix B of 
\cite{Hwang:2012jh},
we can rewrite it into the following form
\begin{align}
&Z_{V}^{\{i_\al\}}
=
\sum_{m_1,\cdots,m_N=0}^\infty 
\prod_{\al=1}^N
\left((-1)^Nq^{N_f}z_\al\right)^{m_\al}q^{-km_\al^2/2}
\nonumber\\
&\times
\prod_{\al=1}^{N}\prod_{l=1}^{m_\al} \,
\frac{
1}{
\prod_{\be=1}^N 2\sinh\pi b(E_{i_\al i_\be}-ilb)
\prod_{j=1, \not\in\{i_\al\}}^{2N_f}2\sinh \pi b (E_{j i_\al}+ilb)
},
\end{align}
and this is precisely the K-theoretic uplift of the vortex partition function
for $U(N)$ theory with $2N_f-N$ fundamentals \cite{Shadchin:2006yz} .




\begin{thebibliography}{99}


\bibitem{Pestun:2007rz} 
  V.~Pestun,
  ``Localization of gauge theory on a four-sphere and supersymmetric Wilson loops,''
  Commun.\ Math.\ Phys.\  {\bf 313}, 71 (2012)
  [arXiv:0712.2824 [hep-th]].
  


  
  
  \bibitem{Festuccia:2011ws} 
  G.~Festuccia and N.~Seiberg,
  ``Rigid Supersymmetric Theories in Curved Superspace,''
  JHEP {\bf 1106}, 114 (2011)
  [arXiv:1105.0689 [hep-th]].


\bibitem{Kapustin:2010xq} 
  A.~Kapustin, B.~Willett and I.~Yaakov,
  ``Nonperturbative Tests of Three-Dimensional Dualities,''
  JHEP {\bf 1010}, 013 (2010)
  [arXiv:1003.5694 [hep-th]].

\bibitem{Kapustin:2010mh} 
  A.~Kapustin, B.~Willett and I.~Yaakov,
  ``Tests of Seiberg-like Duality in Three Dimensions,''
  arXiv:1012.4021 [hep-th].
  
  \bibitem{Drukker:2010nc} 
  N.~Drukker, M.~Marino and P.~Putrov,
  ``From weak to strong coupling in ABJM theory,''
  Commun.\ Math.\ Phys.\  {\bf 306}, 511 (2011)
  [arXiv:1007.3837 [hep-th]].

  

\bibitem{Jafferis:2010un} 
  D.~L.~Jafferis,
  ``The Exact Superconformal R-Symmetry Extremizes Z,''
  JHEP {\bf 1205}, 159 (2012)
  [arXiv:1012.3210 [hep-th]].
  
  
 \bibitem{Hama:2010av} 
  N.~Hama, K.~Hosomichi and S.~Lee,
  ``Notes on SUSY Gauge Theories on Three-Sphere,''
  JHEP {\bf 1103}, 127 (2011)
  [arXiv:1012.3512 [hep-th]].


  
  \bibitem{Benvenuti:2011ga} 
  S.~Benvenuti and S.~Pasquetti,
  ``3D-partition functions on the sphere: exact evaluation and mirror symmetry,''
  JHEP {\bf 1205}, 099 (2012)
  [arXiv:1105.2551 [hep-th]].


\bibitem{Pasquetti:2011fj} 
  S.~Pasquetti,
  ``Factorisation of N = 2 Theories on the Squashed 3-Sphere,''
  JHEP {\bf 1204}, 120 (2012)
  [arXiv:1111.6905 [hep-th]].

  
  
  
  \bibitem{Shadchin:2006yz} 
  S.~Shadchin,
  ``On F-term contribution to effective action,''
  JHEP {\bf 0708}, 052 (2007)
  [hep-th/0611278].

  \bibitem{Dimofte:2010tz} 
  T.~Dimofte, S.~Gukov and L.~Hollands,
  ``Vortex Counting and Lagrangian 3-manifolds,''
  Lett.\ Math.\ Phys.\  {\bf 98}, 225 (2011)
  [arXiv:1006.0977 [hep-th]].


  
  \bibitem{Doroud:2012xw}
    N.~Doroud, J.~Gomis, B.~Le Floch and S.~Lee,
  ``Exact Results in D=2 Supersymmetric Gauge Theories,''
  arXiv:1206.2606 [hep-th].
  
\bibitem{Benini:2012ui} 
  F.~Benini and S.~Cremonesi,
  ``Partition functions of N=(2,2) gauge theories on $S^2$ and vortices,''
  arXiv:1206.2356 [hep-th].
  
  
  
  


\bibitem{Beem:2012mb} 
  C.~Beem, T.~Dimofte and S.~Pasquetti,
  ``Holomorphic Blocks in Three Dimensions,''
  arXiv:1211.1986 [hep-th].

   



 

\bibitem{Hama:2011ea} 
  N.~Hama, K.~Hosomichi and S.~Lee,
  ``SUSY Gauge Theories on Squashed Three-Spheres,''
  JHEP {\bf 1105}, 014 (2011)
  [arXiv:1102.4716 [hep-th]].  
  
  \bibitem{Imamura:2011wg} 
  Y.~Imamura and D.~Yokoyama,
  ``N=2 supersymmetric theories on squashed three-sphere,''
  Phys.\ Rev.\ D {\bf 85}, 025015 (2012)
  [arXiv:1109.4734 [hep-th]].

 
 
 \bibitem{Kim:2009wb} 
  S.~Kim,
  ``The Complete superconformal index for N=6 Chern-Simons theory,''
  Nucl.\ Phys.\ B {\bf 821}, 241 (2009)
  [Erratum-ibid.\ B {\bf 864}, 884 (2012)]
  [arXiv:0903.4172 [hep-th]].
 
 \bibitem{Imamura:2011su} 
  Y.~Imamura and S.~Yokoyama,
  ``Index for three dimensional superconformal field theories with general R-charge assignments,''
  JHEP {\bf 1104}, 007 (2011)
  [arXiv:1101.0557 [hep-th]].
  
  
  
  
\bibitem{Iqbal:2012xm} 
  A.~Iqbal and C.~Vafa,
  ``BPS Degeneracies and Superconformal Index in Diverse Dimensions,''
  arXiv:1210.3605 [hep-th].
  
  
    
\bibitem{Dimofte:2011py} 
  T.~Dimofte, D.~Gaiotto and S.~Gukov,
  ``3-Manifolds and 3d Indices,''
  arXiv:1112.5179 [hep-th].
  
  
  \bibitem{Hwang:2012jh} 
  C.~Hwang, H.~-C.~Kim and J.~Park,
  ``Factorization of the 3d superconformal index,''
  arXiv:1211.6023 [hep-th].
  

  
  
     
  
  \bibitem{Cecotti:2010fi} 
  S.~Cecotti, A.~Neitzke and C.~Vafa,
  ``R-Twisting and 4d/2d Correspondences,''
  arXiv:1006.3435 [hep-th].
  
     

  
  
  \bibitem{Marino:2011eh} 
  M.~Marino and P.~Putrov,
  ``ABJM theory as a Fermi gas,''
  J.\ Stat.\ Mech.\  {\bf 1203}, P03001 (2012)
  [arXiv:1110.4066 [hep-th]].
  
  





 



  
  
  
  \bibitem{Aganagic:2003db} 
  M.~Aganagic, A.~Klemm, M.~Marino and C.~Vafa,
  ``The Topological vertex,''
  Commun.\ Math.\ Phys.\  {\bf 254}, 425 (2005)
  [hep-th/0305132].
  
  \bibitem{Iqbal:2004ne} 
  A.~Iqbal and A.~-K.~Kashani-Poor,
  ``The Vertex on a strip,''
  Adv.\ Theor.\ Math.\ Phys.\  {\bf 10}, 317 (2006)
  [hep-th/0410174].
  
  
  
\bibitem{Bonelli:2011fq} 
G.~Bonelli, A.~Tanzini and J.~Zhao,
``Vertices, Vortices and Interacting Surface Operators,''
JHEP {\bf 1206}, 178 (2012)
[arXiv:1102.0184 [hep-th]].



  
  
  \bibitem{Alday:2009fs} 
  L.~F.~Alday, D.~Gaiotto, S.~Gukov, Y.~Tachikawa and H.~Verlinde,
  ``Loop and surface operators in N=2 gauge theory and Liouville modular geometry,''
  JHEP {\bf 1001}, 113 (2010)
  [arXiv:0909.0945 [hep-th]].

  
  \bibitem{Kozcaz:2010af} 
  C.~Kozcaz, S.~Pasquetti and N.~Wyllard,
  ``A \& B model approaches to surface operators and Toda theories,''
  JHEP {\bf 1008}, 042 (2010)
 [arXiv:1004.2025 [hep-th]].
 
 \bibitem{Taki:2010bj} 
 M.~Taki,
 ``Surface Operator, Bubbling Calabi-Yau and AGT Relation,''
JHEP {\bf 1107}, 047 (2011)
[arXiv:1007.2524 [hep-th]].

\bibitem{Awata:2010bz} 
H.~Awata, H.~Fuji, H.~Kanno, M.~Manabe and Y.~Yamada,
``Localization with a Surface Operator, Irregular Conformal Blocks and Open Topological String,''
arXiv:1008.0574 [hep-th].









  \bibitem{Bonelli:2011wx} 
  G.~Bonelli, A.~Tanzini and J.~Zhao,
  ``The Liouville side of the Vortex,''
  JHEP {\bf 1109}, 096 (2011)
  [arXiv:1107.2787 [hep-th]].

\bibitem{Chen:2013dda} 
  H.~-Y.~Chen and A.~Sinkovics,
  ``On Integrable Structure and Geometric Transition in Supersymmetric Gauge Theories,''
  arXiv:1303.4237 [hep-th].





\bibitem{Nieri:2013yra} 
F.~Nieri, S.~Pasquetti and F.~Passerini,
``3d \& 5d gauge theory partition functions as q-deformed CFT correlators,''
arXiv:1303.2626 [hep-th].

\bibitem{Bao:2011rc} 
L.~Bao, E.~Pomoni, M.~Taki and F.~Yagi,
``M5-Branes, Toric Diagrams and Gauge Theory Duality,''
JHEP {\bf 1204}, 105 (2012)
[arXiv:1112.5228 [hep-th]].
  
 
 
\bibitem{Shintani} 
  T.~Shintani, 
  ``On a Kronecker limit formula for real quadratic fields,''
   J.\ Fac.\ Sci.\ Univ.\ Tokyo\ Sect.\ 1A Math.  {\bf 24} (1977) 167

 \bibitem{Barnes} 
  E.~W.~Barnes,
  `` Theory of the double gamma function,"
   Phil.\ Trans.\ Roy.\ Soc.\ A {\bf 196} (1901) 265-388
   
   
   
   \bibitem{Eguchi:2003sj} 
  T.~Eguchi and H.~Kanno,
  ``Topological strings and Nekrasov's formulas,''
  JHEP {\bf 0312}, 006 (2003)
  [hep-th/0310235].
\end{thebibliography}
\end{document}